\documentclass[twocolumn]{aastex631}
\usepackage{amsmath}

\begin{document}

\title{Persistent Homology analysis for solar magnetograms}

\author{P. Santamarina Guerrero}
\affiliation{Instituto de Astrofísica de Andalucía (CSIC), Apdo. de Correos 3004, E-18080 Granada, Spain}
\altaffiliation{Spanish Space Solar Physics Consortium (S$^3$PC)}

\author{ Yukio Katsukawa}
\affiliation{ National Astronomical Observatory of Japan, 2-21-1 Osawa, Mitaka, Tokyo 181-8588, Japan}

\author{Shin  Toriumi}
\affiliation{Institute of Space and Astronautical Science, Japan Aerospace Exploration Agency, 3-1-1, Yoshinodai, Chuo-ku, Sagamihara, Kanagawa 252-5210, Japan}

\author{D. Orozco Suárez}
\affiliation{Instituto de Astrofísica de Andalucía (CSIC), Apdo. de Correos 3004, E-18080 Granada, Spain}
\altaffiliation{Spanish Space Solar Physics Consortium (S$^3$PC)}

\begin{abstract}

Understanding the magnetic fields of the Sun is essential for unraveling the underlying mechanisms driving solar activity. Integrating topological data analysis techniques into these investigations can provide valuable insights into the intricate structures of magnetic fields, enhancing our comprehension of solar activity and its implications. In this study, we explore what persistent homology can offer in the analysis of solar magnetograms, with the objective of introducing a novel tool that will serve as the foundation for further studies of magnetic structures at the solar surface. By combining various filtration methods of the persistent homology analysis, we conduct an analysis of solar magnetograms that captures the broad magnetic scene, involving a mixture of positive and negative polarities. This analysis is applied to observations of both quiet Sun and active regions, taken with Hinode/SOT and SDO/HMI, respectively. Our primary focus is on analyzing the properties of the spatial structures and features of the magnetic fields identified through these techniques. The results show that persistent diagrams can encode the spatial structural complexity of the magnetic flux of active regions by identifying the isolated, connected, and interacting features. They facilitate the classification of active regions based on their morphology and the detection and quantification of interacting structures of opposing polarities, such as $\delta$-spots. The small-scale events in the quiet Sun, such as magnetic flux cancellation and emergence, are also revealed in persistent diagrams and can be studied by observing the evolution of the plots and tracking the relevant features.

\end{abstract}


\section{Introduction} \label{sec:intro}

The ability to encode and simplify all information about the shape and distribution of data has made Topological Data Analysis (TDA) one of the most relevant fields in state-of-the-art data analysis. In recent years, we have witnessed a rise of studies based on TDA techniques in many fields of science, such as biomedicine \citep{brain-PH}, atomic physics \citep{atomic}, image recognition \citep{Image-recognition} or cosmology \citep{cosmology}, among many others.

Among the numerous techniques of TDA, persistent homology is arguably the most widely used approach for studying real data. By examining the persistence of topological features, persistent homology can identify significant structures present at different scales, and at the same time, its performance is very robust against noisy and/or incomplete data \citep{PH_noisy}. Furthermore, persistent homology provides a straightforward and intuitive way for the visualization of the results. This simplifies the interpretation of the results, while also serving as a good descriptor of the data's topological properties, therefore making it a suitable input for machine learning algorithms.

The application of these techniques in solar observations presents a promising approach to understanding the complex structures and dynamics of the Sun's behavior. Specifically, the analysis of the solar magnetic field using magnetograms is particularly well-suited for the application of these methodologies, given the intricate and multi-scale nature of the magnetic structures. Solar magnetograms provide a visual and quantitative representation of the magnetic field in the photosphere and are one of the fundamental tools for the study of our star. The magnetic activity of the Sun is very diverse, from the quieter events occurring in the quiet Sun to the more violent and extreme events like solar flares and coronal mass ejections (CMEs) in active regions. In this sense, magnetograms are very useful as they enable us to study all these events through magnetic field measurements.

Numerous studies utilize magnetograms to investigate the behavior of solar magnetic fields. The intricate nature of the magnetic structures has led to the development of various techniques, each tailored to focus on distinct properties of the magnetic field. One of these techniques is the study of the power spectrum of magnetograms through Fourier transforms, as employed in numerous works: in \cite{power_spectrum1}, where they attempt to establish a correlation between the magnetogram power spectrum and flare production; in \cite{power_spectrum_2}, where they use the magnetogram power spectrum to study the quiet-sun turbulence; in \cite{power_spectrum_3}, where they analyzed the power spectrum of different physical quantities and study their dependence with the total magnetic flux; or in  \cite{power_spectrum_4}, where they tried to reproduce the magnetograms power spectrum through simulations, among other instances. 

Different approaches are also common. Some examples of alternative methodologies can be found in: \cite{intermitency}, where they study the intermittency and multifractality of the magnetic structures and their relation with flaring activity; in \cite{conectivity}, where they study the magnetic connectivity to define a criteria for the distinction of flaring and nonflaring regions; or in \cite{Milan}, where they analyze long time series of magnetograms with high cadence and spatial resolution to calculate the number of field appearances and cancellations, as well as their interactions, to determine the net magnetic fluxes on the Sun's surface; among many other approaches in the field.

The increasing volume of data generated by modern instruments highlights the growing importance of data analysis techniques. Many studies have directed their efforts towards the development of automatic feature detection and tracking algorithms for solar magnetograms. Prominent examples of widely employed approaches for Quiet Sun studies include SWAMIS\footnote{The Southwest Automatic Magnetic Identification Suite} \citep{swamis_yafta}, as employed, for example, in \cite{swamis_example}, where they employ the code to track the magnetic elements and study the flux dispersal in the Quiet Sun. Another example is YAFTA\footnote{Yet Another Feature Tracking Algorithm} \citep{yafta}, employed in \cite{yafta_example},  where they track the proper motion of magnetic elements of the Quiet Sun to study the dynamics of supergranular flows. Concerning active regions, there have been numerous works on the matter of classification and detection methods, from the well-known, and classical approach of the Mount-Wilson classification \citep{hale}, to more recent contributions, such as the SHARP\footnote{Spaceweather HMI Active Region Patch} tool \citep{sharp}, that has emerged as one of the most prominent algorithms for this purpose.

Although these studies provide valuable insight into the processes occurring in the photosphere and the interrelations of the solar magnetic field with other solar phenomena, the underlying governing laws remain highly complex and challenging to fully ascertain. The integration of TDA techniques into these analyses has the potential to offer a previously unexplored perspective on these phenomena that complements the current knowledge. Persistent homology algorithms share similar methodologies (such as image thresholding) with other feature detection/tracking codes like SWAMIS or YAFTA. However, unlike these codes, persistent homology provides topological information about the detected features and employs it to discern between different types of structures. This includes information on the connectivity to neighboring features, the shape of the feature, and the presence or absence of holes in the magnetic feature. All this information allows persistent homology to distinguish between different types of magnetic features based on their topological properties, and to identify and track the presence of a particular type of magnetic structure. In addition, since these algorithms can identify the pixels that make up each topological feature, they can be used to outline magnetic elements and facilitate conventional calculations such as determining the size or flux of magnetic structures.

In particular, topological techniques can be particularly useful for studies related to solar flares. It is well known that active regions exhibiting intricate structures are linked to the occurrence of solar flares. Three critical factors establish a connection between the characteristics of active regions and flare production: surface area, magnetic complexity, and rapid temporal evolution \citep{flare_LR}. While the first factor is straightforward to measure (e.g., the sunspot area and total unsigned magnetic flux), persistent homology techniques may enable topological quantification of the latter two, which present greater challenges when utilizing conventional methods. Some studies have already delved into this concept; for example, in \cite{ph_solar_2}, they employ a persistent homology analysis to explore the predictive capabilities of a machine learning model for the forecasting of solar flares based on the topological information extracted from solar magnetograms. However, they did not study the correspondence between the magnetic features and the topological information extracted from persistent homology, which is the main focus of this work.

In this study, we describe the most appropriate approach when employing persistent homology in the specific case of solar magnetograms to capture the broad complexity of the spatial structure. We also analyze the different structures the algorithm can find and show how to identify specific magnetic field behaviors using the tools provided by persistent homology, such as persistent diagrams and images. With this work, our objective is to provide the groundwork for future studies, enabling them to incorporate these techniques in fields of solar physics where the characterization of magnetic field structures is fundamental. In sect.~\ref{Sec: PH} we present the fundamentals of persistent homology and the interpretation of its results. Section ~\ref{sec : data} includes a description of the data used. The application of persistent homology and its analysis is discussed in sect. ~\ref{sec : Analysis}. Lastly, we present some conclusions in section \ref{sec : Conclusion}.

\section{\label{Sec: PH} Persistent Homology}

Persistent homology stands out as a prominent technique within the topological data analysis toolkit, primarily for its capacity to capture the shape and distribution information of a dataset. The algorithm is rooted in the mathematical framework of homology groups. In topology, these groups measure the number of $n$-dimensional holes in a data set, or in other words, the number of connected components for a  $0^{th}$ dimensional analysis, holes or rings for a $1^{st}$ dimensional analysis, spherical voids for the $2^{nd}$ dimensional analysis, and so on. 

The primary objective of persistent homology is not only to compute the homology groups of a given dataset but also to study how they vary at different scales. To achieve this, the input data undergoes a process of division into a series of sequential subspaces, with each subspace encompassing the previous one. This sequential process, known as filtration, begins with a starting subspace comprising a single point from the original dataset. Subsequent subspaces are then constructed by incrementally adding points to the previous subspace until the final subspace includes all points of the original dataset. 

After the filtration process is performed, persistent homology algorithms shift their focus to analyzing the evolution of topological features across the different subspaces. Specifically, they record the filtration value at which a new feature appears, meaning that it is absent in the previous subspace, and when it disappears, meaning that it is no longer present in the following subspaces. These two events are known as the birth and death of a topological feature, respectively. 

In a nutshell, the $n$-dimensional persistent homology of a dataset with a given filtration can be described as the aggregation of all n-dimensional features (homology groups) that were created (birth) and subsequently eliminated (death) during the filtration process \citep{PH_survey_for_ML}.

When applying persistent homology on a greyscale image, our focus lies in filtering the data according to the pixel values. Multiple filtering approaches exist, with the most extended ones being sublevel and superlevel filtrations, both based on the concept of thresholding. In these filtrations, the image is cropped to a specific value, forming a subspace that includes all pixels with values higher than this value in a superlevel filtration, or lower in a sublevel filtration. This cropping value (i.e. the filtration value) is systematically varied from the lowest to the highest values of the image, or vice versa, thus generating a different subspace for each value. As a result, the persistence homology analysis captures and examines the evolution of topological features across different thresholds, enabling insights into the image's structural properties at various scales \citep{PH-Methods-Danielle}.

A more formal way of defining these filtrations can be done by considering an image as a discrete representation of a function $f$, defined over a two-dimensional space $\mathbb{X}$, such that: 
\begin{equation}
    f :  \mathbb{X} \longrightarrow  \mathbb{R} \ \ .
\end{equation}
Let $\mathbb{S} _\phi$ be the subspace of $\mathbb{X}$ for a filtration value of $\phi$. In such a case, a filtration can be expressed as:
\begin{equation}
    \mathbb{X}: \mathbb{S}_{\phi_0} \subset \mathbb{S}_{\phi_1} \subset \mathbb{S}_{\phi_2}\subset... \subset \mathbb{X}\ \ .
\end{equation}
With this formulation, a topological feature with birth-death coordinates:
\begin{equation}
(B, D) = (\phi_ I, \phi _ {II}) \ \ ,
\end{equation} 
corresponds to a feature that appears for the first time during the filtration process at the subspace $\mathbb{S}_{\phi _ I}$, and \textit{persists} until the subspace $\mathbb{S}_{\phi _ {II}}$, where it ceases to exist.

In a sublevel filtration, each subspace can be expressed as:
\begin{equation}
    \mathbb{S} _ \phi = f ^{-1} \left( ( -\infty, \phi ] \right) \ \ ,
    \label{eq: sublevel}
\end{equation}
where $\phi_0$ is selected as the lowest value for any given pixel and its value is increased until the subspace includes all pixels. On the contrary, in a superlevel filtration, the subspaces can be expressed as:
\begin{equation}
    \mathbb{S} _ \phi = f ^{-1} \left( [ \phi, \infty ) \right)\ \ , 
    \label{eq : superlevel}
\end{equation}
where $\phi_0$ is selected as the highest value for any given pixel and its value is decreased along the filtration.

Various methods exist for representing the information derived from a persistent homology analysis, including Betti numbers, persistence bars, and persistent diagrams (PDs) (\citealt{PD_stab}, \citealt{persistence-bars-betti}), among many others. For this study, we will utilize the PDs as our chosen approach due to their straightforward interpretation and extended use. A $n$-dimensional PD is a multiset of Birth-Death pairs, ($B _ i, D _ j$), with multiplicity $k$, where each pair measures the number ($k$) of $n$-dimensional components that have been born at the filtration subspace $\mathbb{X}_i$ and died in $\mathbb{X}_j$, that is usually represented in a 2D scatter plot.

\begin{figure}
    \centering
     \includegraphics[width=8cm]{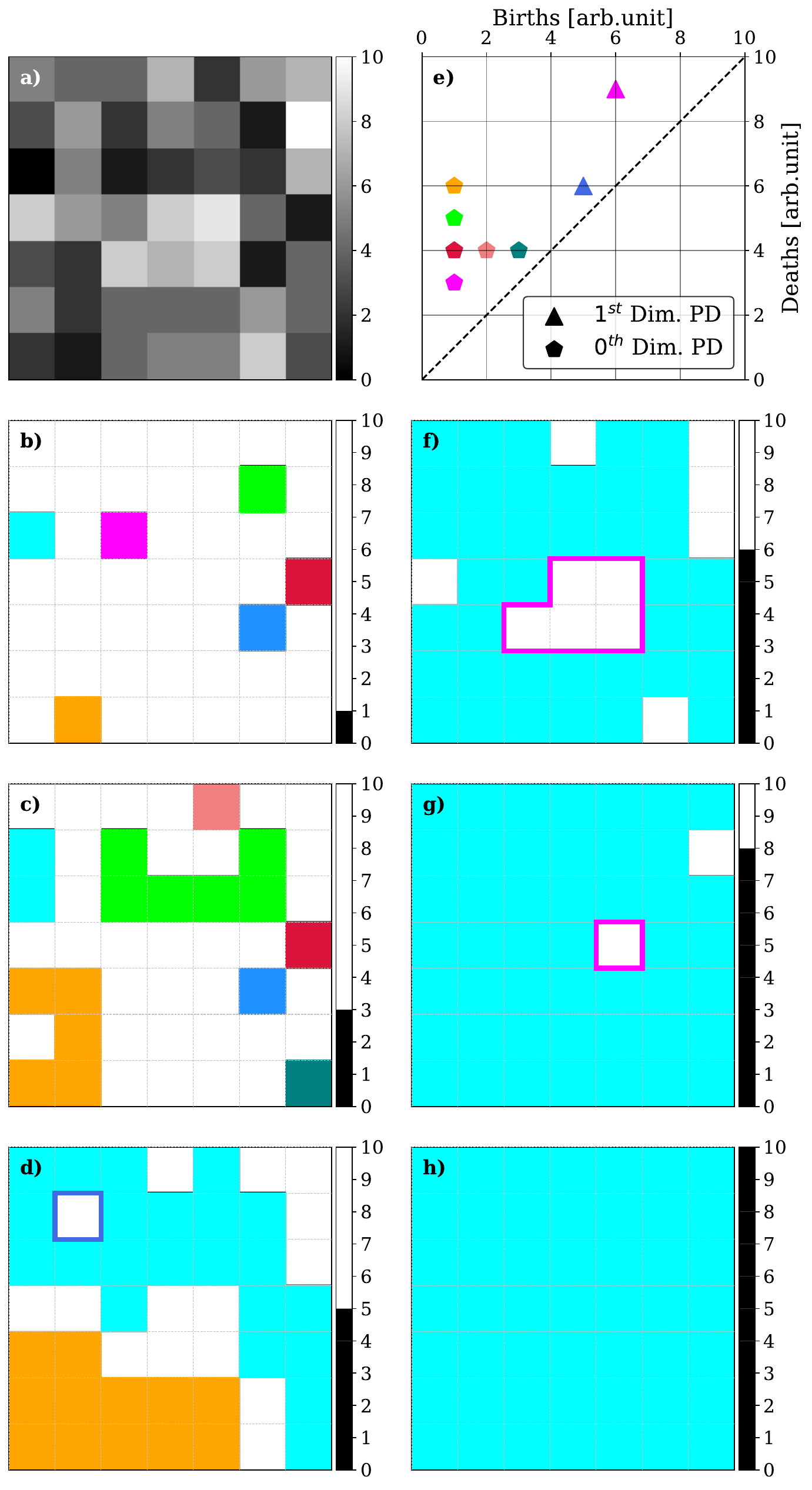}
    \caption{Sublevel filtration of a greyscale image and PDs of the $0^{th}$ and $1^{st}$ dimensions. Panel a) shows the input data. In panels b), c), d), f), g), and h) different snapshots of the filtration process are shown. The value of the filtration parameter, $\phi$ is shown in the color bar at the right of each image. Only pixels with a value lower than the filtration value (colored pixels) belong to the subspace shown in each snapshot. Different homology groups are represented with different colors at each snapshot. For connected components ($0^{th}$ dimensional homology groups) the whole pixel is shown with the corresponding color. For rings, ($1^{st}$ dimensional homology groups), only the border of each hole is colored. In panel e) the PDs of both dimensions are shown. The color of each point in the diagram is the same as the one used to plot the corresponding topological feature in other panels. An animation depicting the whole filtration process is also provided as part of the article.}
   \label{fig: PD Example Images}
\end{figure}

The process of generating a PD of a greyscale image is as follows. We start by selecting the filtration direction (sublevel or superlevel) and the dimension of the analysis (either 0 or 1). We initialize a threshold as the highest or lowest value from the image, depending on the choice of filtration. We then perform the filtration by systematically adjusting the threshold and creating a binary image for each threshold. This process divides the image into two sets: pixels with values above the threshold and pixels with values below it. The choice of filtering determines which of the two sets makes up the subspace. We then look for the existing topological features within each of these subdivisions. The specific process by which these features are identified is detailed in the next paragraph, where the structures corresponding to both dimensions are illustrated using the example shown in Fig.~\ref{fig: PD Example Images}. We repeat this process until the threshold reaches the opposite limit to that from which it started. Along this process, we follow the appearance, merging, and disappearance of connected components. When two components merge, the longer-lived one (\textit{i.e}. the first to appear along the filtration process) absorbs the younger one, thus resulting in the death of the second \citep{eldest}. We determine the birth and death for each component based on the thresholds at which these events occur.  Finally, we construct a scatter plot where the horizontal axis represents the birth values and the vertical axis represents the death values. Each point on this plot corresponds to a persistence point, whose coordinates reveal the scales at which the corresponding topological feature is present.

An example of this process with a sublevel filtration is shown in Fig.~\ref{fig: PD Example Images}. Panel a) displays the input data, panels b), c), d), f), g), and h) show some of the key steps of the filtration process, and, lastly, panel e) displays the PD for a $0^{th}$ and $1^{st}$ dimensional analyses. These plots illustrate how connected components and rings are born and then die as we increase the filtration level. As these components (shown in different colors) increase in size and come into contact with other components, one absorbs the other, thus resulting in the death of the second. This phenomenon is shown in panels b) to d), where we observe the progression of the components until only the blue and orange connected components remain. Additionally, the diagrams also reveal the appearance of two rings in the data (panels f) and g)). These rings are found when pixels that do not belong to the subspace are surrounded by a connected component, and die when those pixels are included in the component as the threshold increases (blue ring in panel f)). Finally, the PD (panel e)) displays the birth and death values (i.e. the filtration value) of all the features, of dimensions 0 and 1, that have been identified (birth) and subsequently eliminated (death) throughout the filtration process. 

\subsection{Persistent images}

The PD displayed in Fig.~\ref{fig: PD Example Images} contains only a limited number of points due to the simplicity of the input image. However, when analyzing real data, these diagrams can consist of hundreds or even thousands of birth-death pairs with high multiplicities, simply due to the size of the images. Additionally, features not only representing the genuine behavior of the data but also reflecting the distribution of noise appear on the diagrams. To address this complexity, several strategies have been developed to simplify the information from PDs, such as persistence curves \citep{persistence-curves}, persistence landscapes \citep{Persistence-landscapes}, or persistence images (PI) \citep{Peristence-images}. In this study, we will focus on the latter, due to its noise filtering capabilities and because the representation of the results remains in a Birth-Death diagram, allowing for easy interpretation of the results, similar to a persistence diagram.

PIs are a condensed form of a persistence diagram, offering a concise and easy-to-understand representation of its topological features.  They capture the spatial distribution and persistence information of these features, allowing for the enhancement of the most relevant ones and filtering of the others. A PI is constructed using the concept of `persistence'. Each topological feature, represented by a point in a PD, has a persistence, $\pi$, of:
\begin{equation}
    \pi = D - B \ \ ,
    \label{eq: lifetime}
\end{equation}
where $(B, D)$ are the corresponding birth-death coordinates in the diagram. A feature with a large persistence is present at different scales in the data and therefore is more likely to represent the real behavior of the data. On the contrary, short-lived features are typically associated with the noise distribution and usually do not provide much information about the data.

When constructing a PI, a weighting function, $\omega (\pi)$, is employed to assign weights to each point in the diagram, ensuring that longer-lived features have greater weights than shorter-lived ones. There are multiple choices for the shape of the weighting function, which are entirely dependent on the aims of the study and data type. The simplest example is often a linear or power-law relation ($\omega (\pi) = a \pi ^b$), where $a$ and $b$ can be tuned to assign progressively higher weights to higher persistencies, thus focusing the study on the longer-lived components. On the other hand, if the objective is to filter out noise while assigning similar weights to all non-noise points so that all points have a similar relevance in the analysis, the chosen function is usually an arc-tangent.

The PI is then generated by dividing the persistence diagram plane into a grid with a desired resolution. Within each grid region (or pixel), the weighted features of the diagram within the region are added up using a kernel density estimation. The kernel function, $K (z)$, can be tuned to suit the nature and objectives of the analysis, with Gaussian functions being the most common approach. 

\begin{figure}
    \centering
    \includegraphics[width=9cm]{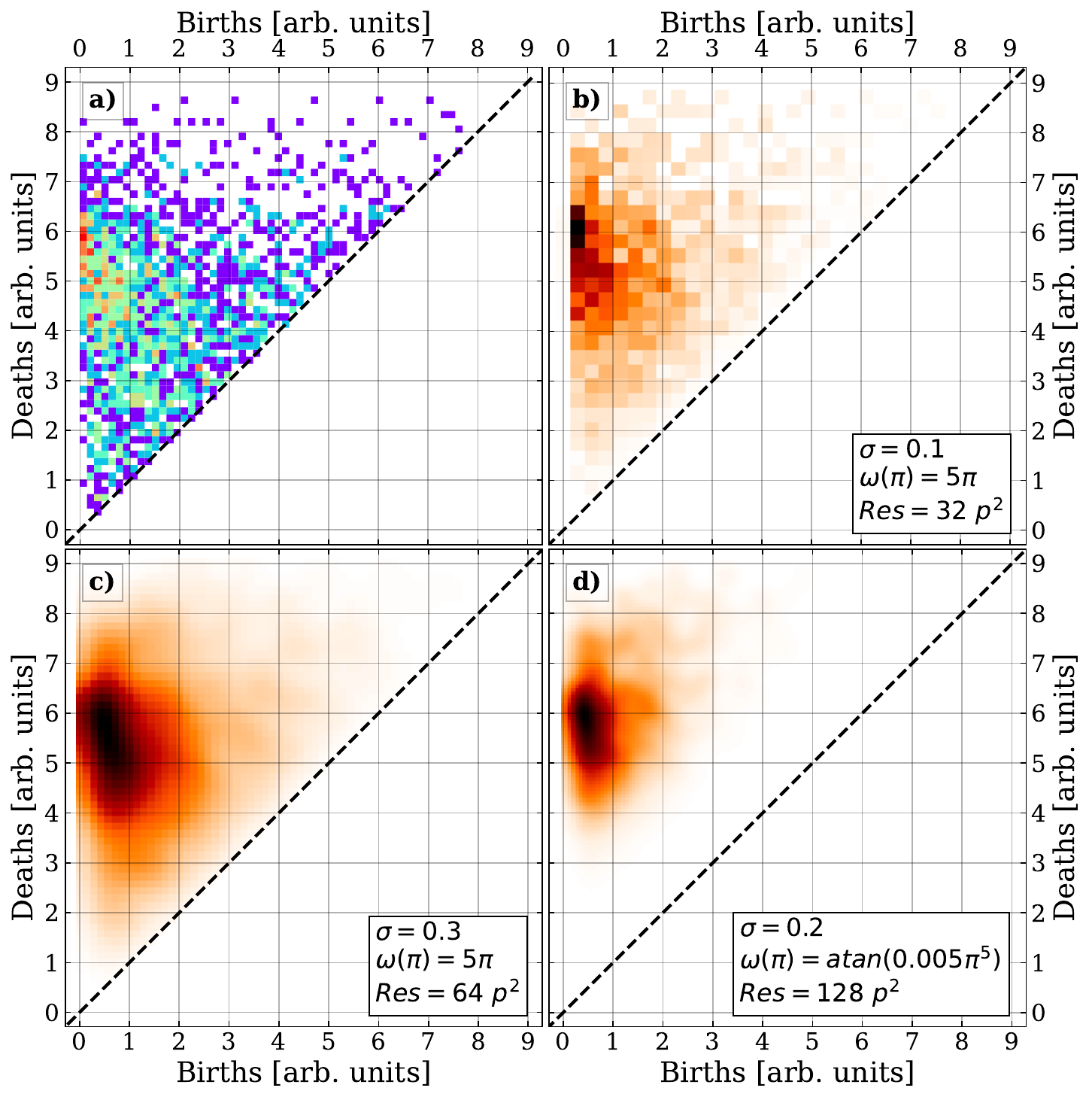}
    \caption{Panel a) shows an example of a persistence diagram as a 2D histogram. The color of each bin represents its multiplicity, with the red spots corresponding to higher values. Panels b), c), and d) show three different examples of PIs. The three parameters given in the legends of the PIs are: the standard deviation ($\sigma$) of the Gaussian kernel ($K (z)$), the weighting function, and the resolution of the image PI in pixels (p).}
   \label{fig: PI Example}
\end{figure}

The resultant PI is a 2D matrix, wherein each pixel corresponds to a specific area in the persistence diagram, and its value represents the cumulative weight of the topological features found within that area. In Figure \ref{fig: PI Example}, three examples of PI (panels b), c), and d)) are presented for the same persistence diagram (panel a)), where distinct choices of resolution, kernel function, and weighting function have been applied to each image.

All the PDs, PIs, and the rest of the analysis tools presented in this work, have been computed using the Homcloud python package \citep{Homcloud}.

\section{\label{sec : data}Data}

In this work, we study the results of applying persistent homology to different regimes of solar activity by applying the analysis to both quiet Sun and active region magnetograms.

\subsection{Quiet Sun observations}

The study of quiet Sun regions requires high magnetic spatial and temporal resolutions and sensitivities to be able to capture the small-scale evolution of the magnetic structures due to their weak signals and short time scales \citep{quiet-sun}. For this reason, we employ observations taken by the Solar Optical Telescope (SOT) \citep{SOT} aboard the \textit{Hinode} satellite \citep{Hinode}, a space-borne solar observatory. In particular, we employ observations from Hinode's Operation Plan (HOP) 151. These observations consist of long ($\ge 20$ h) and mostly uninterrupted sequences of measurements of the Narrowband Filter Imager of the Na I D1 line at 5896 $\AA$ taken with a cadence of $50-70$ s. The data correction of the selected observation sets has been carried out in \cite{milan1}.

\subsection{Active regions observations}

We employ observations of active regions (ARs) taken by the Helioseismic and Magnetic Imager (HMI; \citealt{hmi1}, \citealt{hmi2}) on board the Solar Dynamics Observatory \citep{SDO}. HMI provides a continuous observation of the Sun where a full-disk magnetogram, as well as Dopplergrams, are provided at all times. The full-disk, uninterrupted observations of HMI make it a very suitable instrument to study the evolution of active regions as the formation and development of active regions can be fully captured. 

We focus the analysis on a series of newly-emerging ARs identified in \cite{Toriumi}. In particular, we employed the 12-minute cadence observations taken during the period from May 2010 to June 2011, which corresponded to a period of low solar activity.

\begin{figure}
    \centering
     \includegraphics[width=8cm]{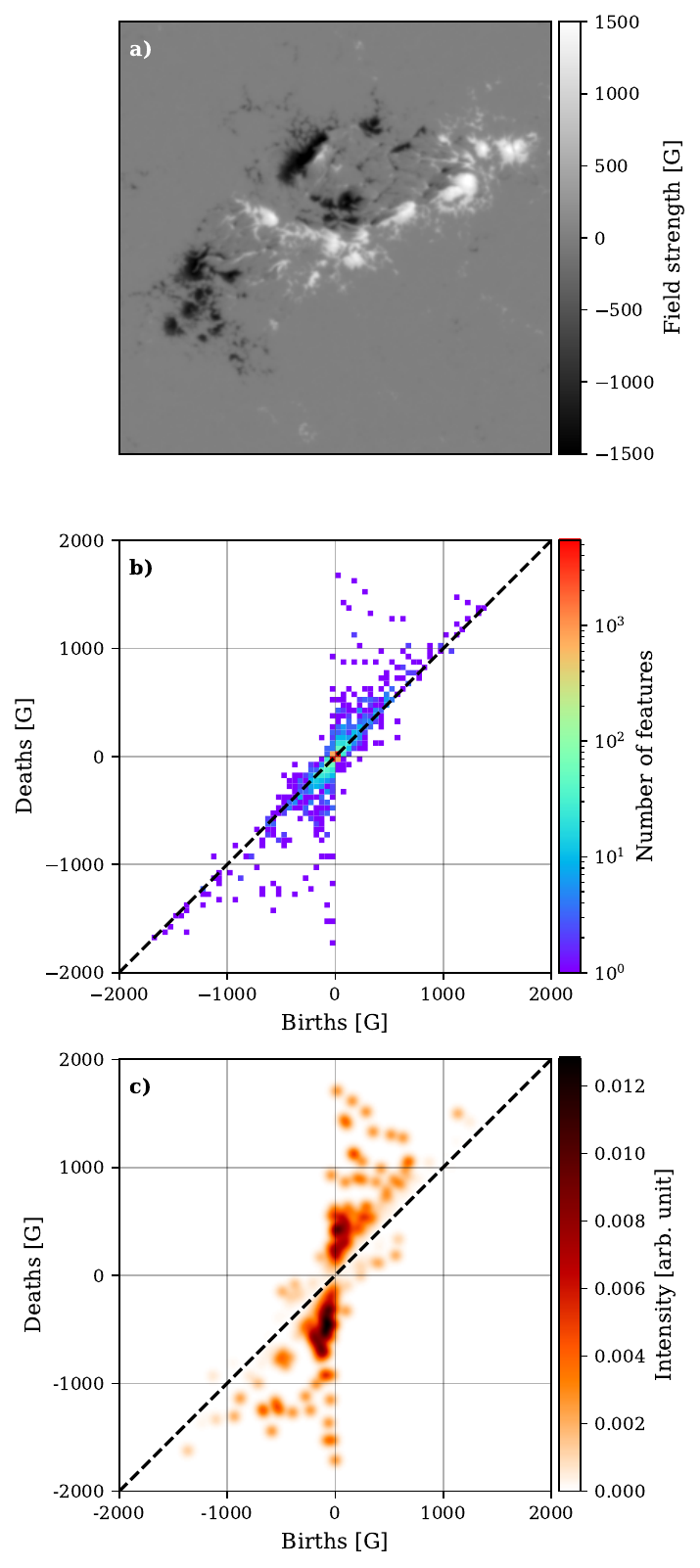}
    \caption{(a) SDO/HMI magnetogram taken on 2011-02-13 depicting an active region (NOAA AR 11158). (b) The corresponding PD combining superlevel and sublevel filtrations. (c) PI generated from the PD in panel b) with the following configuration: Resolution =  1000 pixels$^2$ ($4\ G$ per pixel), weighting function: $\omega (\pi) = \arctan (5\times 10 ^{-8} \pi ^{3})$ and a gaussian kernel with $\sigma = 40\ G$.}
   \label{fig: PD+PI_example}
\end{figure}

\section{\label{sec : Analysis}Analysis and Results}

The application of persistent homology to a specific dataset can vary depending on the aims of the study. Different dimensions of the analysis and various types of filtrations focus on distinct features within the data. It is crucial to have prior knowledge of the expected structures and relevant features to be captured in the analysis in order to determine the appropriate approach. In this section, we aim to outline the most appropriate approach for studying the particular case of solar magnetograms.

The solar magnetograms employed here represent the longitudinal component of the magnetic field on the photosphere and are typically presented as greyscale images, as shown in Figure \ref{fig: PD+PI_example}, panel a). The polarity of the line-of-sight magnetic field is indicated by the sign of each pixel, where positive and negative signals correspond to field lines pointing towards and away from the observer. Applying a single filtration to a greyscale image only displays features corresponding to one polarity (positive or negative) in a PD. However, to conduct a comprehensive study of the magnetic field, both polarities are essential, thus necessitating the use of two separate filtrations with different filtration directions.

\begin{figure}
    \centering
     \resizebox{\hsize}{!}{\includegraphics[width=12cm]{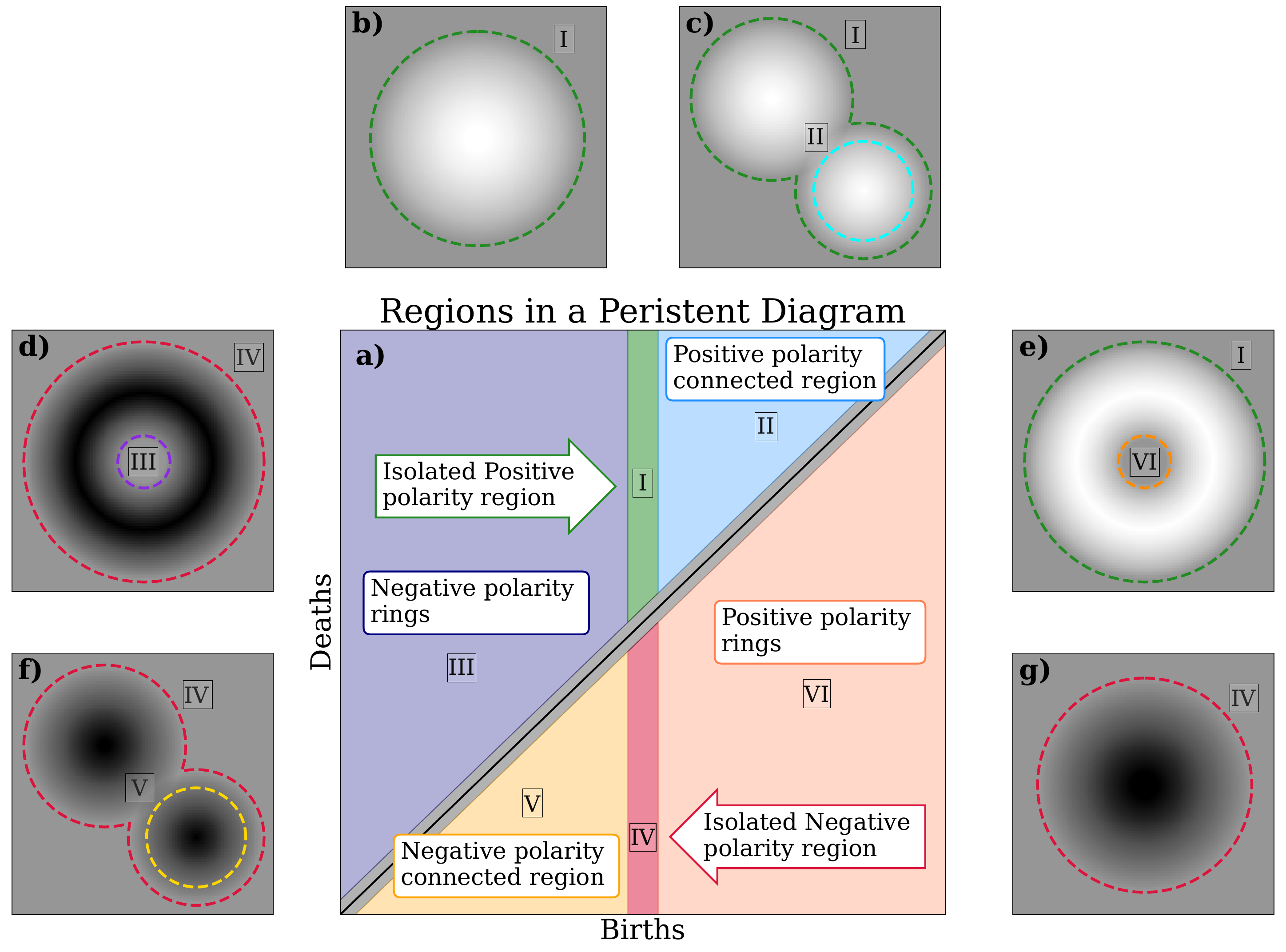}}
    \caption{Schematic representation of the PD and the different regions (panel a)). The magnetic structures corresponding to the topological features found in the different regions are shown in panels b) to g), with each feature identified by a ring with the corresponding color.}
   \label{fig: PD REGIONS}
\end{figure}

We determined that a combination of superlevel and sublevel filtrations with a $1^{st}$-dimensional persistent homology analysis was the most suitable approach for studying solar magnetograms. This choice is based on two main reasons. Firstly, the $1^{st}$ dimensional analysis allows us to identify the most prominent features in a PD, which is not the case in the $0^{th}$ dimensional analysis where the strongest feature does not appear in the diagram as it never dies (see Fig.~\ref{fig: PD Example Images}). Secondly, by combining superlevel and sublevel filtrations, we can display the results of both filtrations in a single diagram. Features corresponding to different filtrations will have persistencies with opposite signs (features found in a superlevel filtration will be born at higher filtration values than their death, resulting in a negative lifespan). This enables us to construct a PD in which all features displayed above the identity line (with positive persistencies) correspond to the sublevel filtration, and those below the line correspond to the superlevel filtration (see panel b) in Fig.~\ref{fig: PD+PI_example}).

The PDs, and consequently the PIs, offer valuable insights into the magnetic structures present in the magnetograms. The location of a topological feature on the diagram is completely determined by the properties of the corresponding magnetic structure. Specifically, this position is influenced by factors such as the maximum intensity of the magnetic field, its proximity to other magnetic structures, and its geometric shape, including the presence of holes or pores within the structure. These characteristics allow us to partition the diagram into distinct regions, where topological features within each region correspond to different types of magnetic structures.

We distinguished between six distinct regions in the diagram. Figure \ref{fig: PD REGIONS} illustrates these regions in panel a) and provides schematic representations of the corresponding magnetic structures in panels b) to g). The regions are as follows: first, topological features located above the identity line (positive persistencies) with birth values close to 0 (region I in the diagram). Features within this region represent isolated magnetic structures of positive polarity, that is, patches of positive magnetic flux fully enclosed by an absence of any magnetic field. The threshold defining what is considered ``close to 0$"$ is determined by the data's properties. To identify isolated structures, we set the threshold as a function of the statistical properties of the background signal (i.e. areas of the magnetogram with little magnetic flux). Specifically, the limits for this region are set as $(-5 \sigma _ {bg}, 5 \sigma _ {bg})$, where $\sigma _ {bg}$ denotes the standard deviation of the background signal found in a $15\times 15$ pixels box devoid of strong magnetic structures.

The second region (II in the diagram) comprises features above the identity line with positive birth values, representing connected structures with positive polarities, that is, positive magnetic field structures in contact with another positive structure but not fully merged. The third region (region III) contains topological features above the identity line with negative birth values, which corresponds to magnetic structures of negative polarity exhibiting a ring-like attribute, namely, structures with pores or holes. These three regions of the diagram have counterparts with negative persistencies. Thus, features associated with isolated structures of negative polarities are found in the region with a birth value close to 0 but below the identity line (region IV), features for connected negative structures are also located below the line but with negative birth values (region V). Lastly, features arising from positive magnetic structures with ring-like attributes are found below the identity line but with positive birth values (region VI).

\subsection{Quiet Sun results}

\begin{figure*}
    \centering
    {\includegraphics[width=18cm]{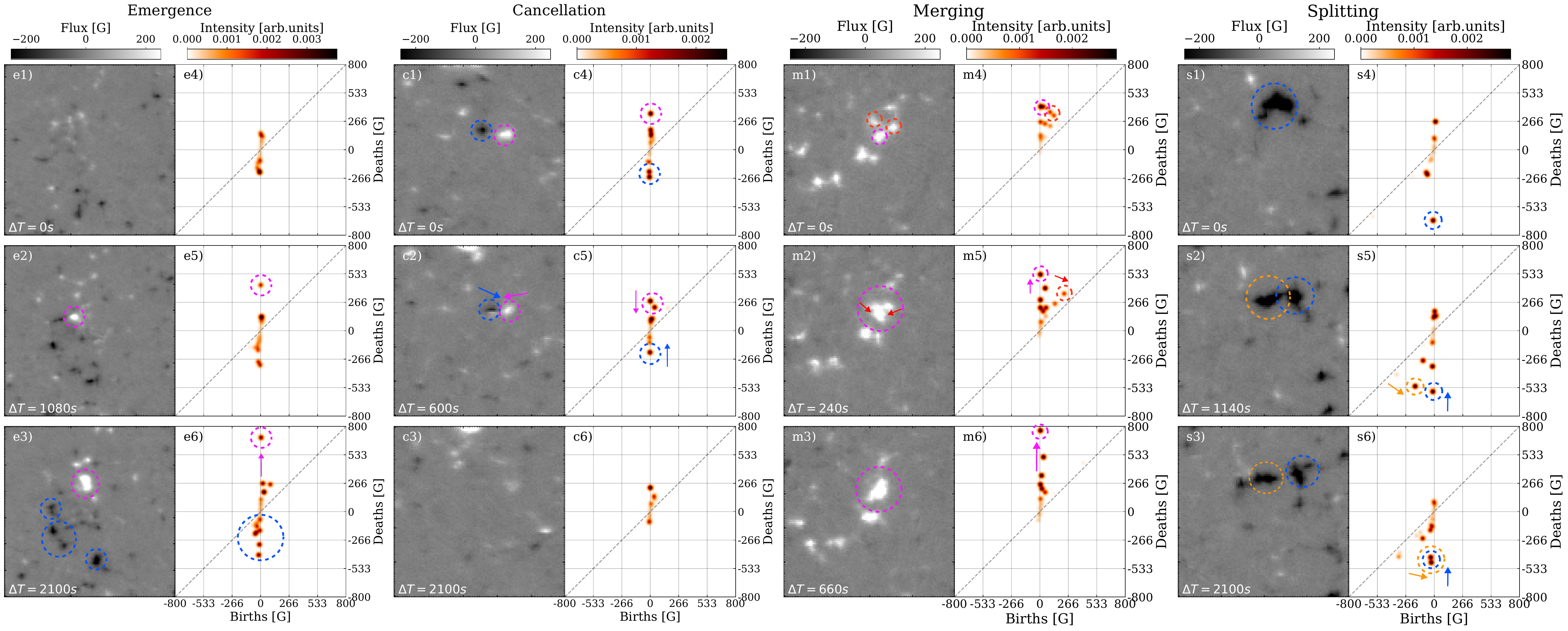}}
    \caption{Examples of flux emergence (left column), flux cancellation (central left column), merging (central right column), and splitting (right column) events in the quiet sun. The time interval shown in the magnetograms is always with respect to the first frame (panel a) for emergence and g) for cancellation). The PIs correspond to the magnetogram to their left. The parameters for their computation are: resolution =  1000 pixels$^2$ (1.6 G per pixel), weighting function: $\omega (\pi) = \arctan (5\times 10 ^{-7} \pi ^{3})$ and a Gaussian kernel with $\sigma = $16 G.}
   \label{fig: Quiet_sun_events}
\end{figure*}

Quiet Sun regions are characterized by the presence of weak, small-scale magnetic field signals that exhibit rapid evolution. This rapid evolution leads to a multitude of signals in a single snapshot that evolve quickly from one frame to another. The small scale and rapid changes of the processes of the quiet Sun make data analysis techniques desirable for their study due to the complexity of such endeavors. 

When examining the evolution of signals across the entire field of view, we observe a dynamic process characterized by numerous regions interacting destructively while new signals emerge throughout the whole region. Despite these continuous changes, the overall structure of the magnetogram appears stable, with consecutive snapshots exhibiting strikingly similar properties. This apparent equilibrium state is also evident when studying the PIs, as consecutive frames show minimal differences in their representations.

Due to the apparent equilibrium state of the overall structure of the magnetograms, it is necessary to narrow down the field to which we apply the analysis. We found that when the number of signals in the studied region is lower, we can observe flux cancellation and emergence events, as well as merging and splitting events, through the changes induced in the persistence diagram. In cancellation events, two regions of magnetic flux with opposite polarities interact destructively, nullifying each other. On the contrary, in flux emergence events, we observe signals of both polarities suddenly emerge from a region with little magnetic flux. In splitting and merging events, only one polarity is involved. Two or more distinct structures of equal polarity merge together in merging events, and a single structure is divided into two or more for splitting events.

In Figure \ref{fig: Quiet_sun_events}, an example of an emergence event is shown in three snapshots through the magnetograms (panels e1 to e3) and their corresponding PIs (panels e4 to e6). When we focus on the second snapshot (panel e2 and e5), we see that a new positive and isolated feature (birth $\sim 0$) that was not present in the previous snapshot has appeared in the PI and stands out from the rest of the signals (highlighted with a pink circle in both magnetogram and PI), while simultaneously, the density of negative polarity features also begins to increase. In the last frame, we see that in the case of positive polarity, the majority of the signal has concentrated in a single structure, as shown by the increase of the death value of the corresponding feature in the PI. A few connected features are also seen, but these are less significant. Meanwhile, the negative polarity signal has been distributed into multiple structures instead of concentrating in a single one, as evidenced by the absence of a prominent feature in the PI and the increased density of connected features (highlighted with blue circles both in the PI and the magnetogram).

\begin{figure*}
    \centering
    {\includegraphics[width=18cm]{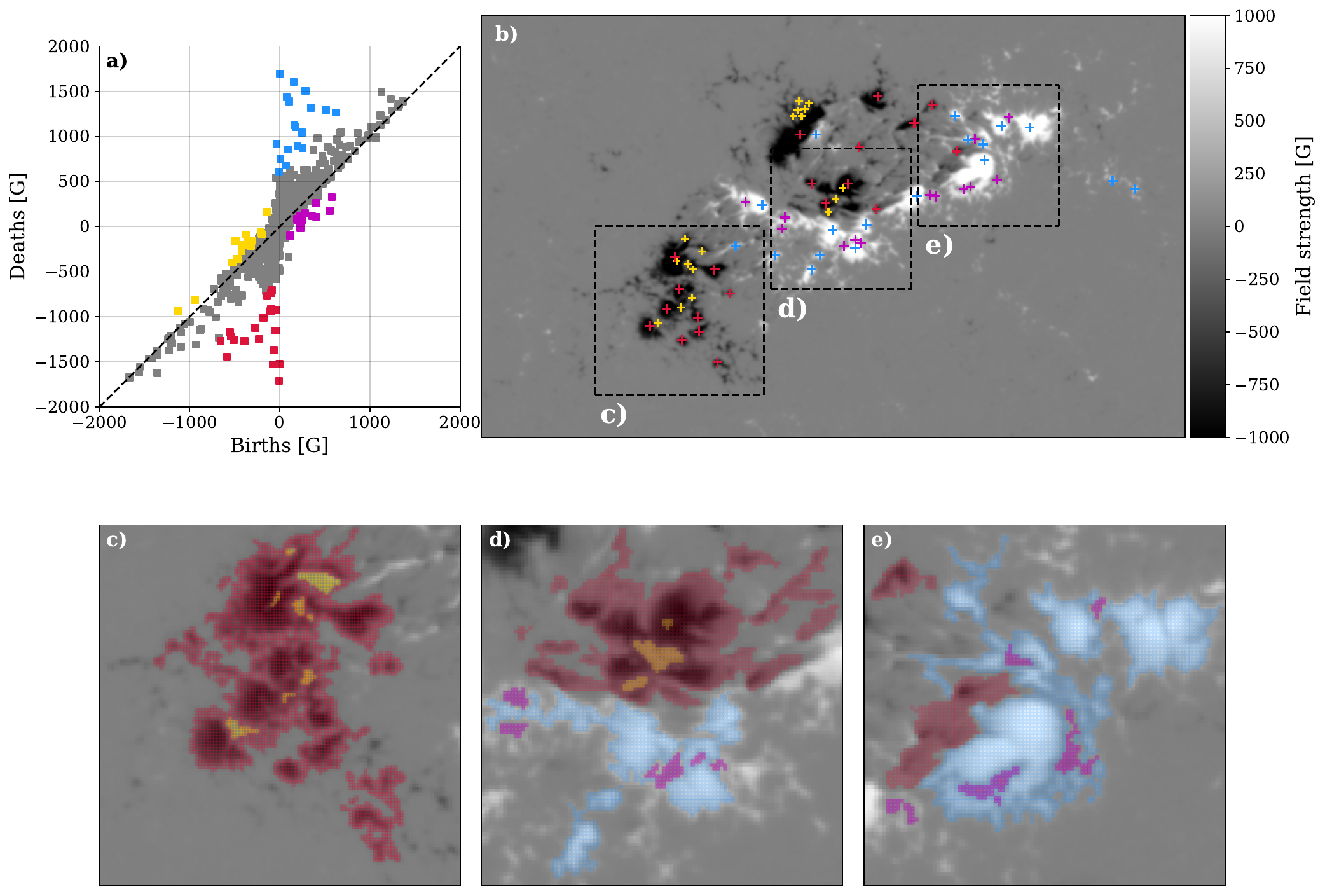}}
    \caption{SDO/HMI magnetogram taken on 2011-02-13 at 06:34 UT depicting an active region (NOAA AR 11158) and the corresponding persistent diagram (panel a)). Panels c) to e) display a zoomed region of the active region, corresponding to the labeled region with the same letter in panel b). Features of different types are shown in different colors in the diagram. The corresponding structures are shown in the same colors in the magnetograms; as colored crosses in panel b) and as a colored transparent overlay displayed over the whole pixel in panels c) to e) to show the extent of the structure.}
   \label{fig:  AR - Complexity}
\end{figure*}

A very similar analysis can be carried out to analyze cancellation events. The evolution of the persistence image is very similar to that of the emergence events but in the opposite direction. Figure \ref{fig: Quiet_sun_events} also shows an example of a flux cancellation event through magnetograms (panels c1 to c3) and PIs (panels c4 to c6). In the beginning, the PI shows the presence of features of opposite polarities. When the corresponding structures approach each other and begin to interact, the magnetic signal starts to decrease, which is observed in the PI as a simultaneous movement of the features towards the center of the diagram. This reduction in signal continues until both features reach the center of the diagram, which corresponds to the moment when they will have completely canceled each other out (panels c3 and c6). 

For the events that only involve one polarity, namely merging and splitting events, the same behavior is seen in the PI for positive and negative features, but on their respective sides of the PI.

An example of a merging event of positive polarity structures is shown in Fig.~\ref{fig: Quiet_sun_events}, in panels m1) to m6). These events start with multiple isolated, or interacting structures (as shown in panels m1 and m4), that are moving towards each other. As the structures cluster, two movements are seen in the PI: firstly, the features corresponding to the structures with the weakest field (red features in panels m2 and m5) move towards the identity line; and secondly, the feature corresponding to the main structure (pink feature in panels m2 and m5) experiences an increase in its absolute death value due to the increase in magnetic flux coming from the rest of the structures. When the structures are fully merged, only a single feature appears in both the magnetogram and PI (pink feature in panels m3 and m6), in the isolated region (region I or IV, depending on the polarity).

This process is reversed for splitting events, as shown in panels s1) to s6) of Fig.~\ref{fig: Quiet_sun_events}. We see how an initially isolated feature in the PI (blue feature in panels s1 and s4) evolves into two (or multiple) features. When the process has started, but the two parts have not yet completely separated, a second feature appears in the diagram in the region corresponding to the connected structures (regions III or V in the diagram), as shown in panels s2) and s5). As the two structures continue to separate, this second feature gradually approaches the region for isolated structures (regions I and IV)  as the magnetic field surrounding it in the magnetogram diminishes (panel s5). Eventually, when both structures are completely separated (i.e. with no signal around them), they will both appear on the diagram as two isolated features with a lower death value compared to the initial one (panels s3 and s5). This reduction occurs because the magnetic flux is distributed between the two structures.

\subsection{Active regions results}

In contrast to the quiet Sun, active regions exhibit a highly complex structure and clear evolution in terms of the overall structure of the magnetic field signals. This complexity, where different features are observed at a wide range of spatial scales, makes these observations an ideal scenario for the application of persistent homology algorithms. Persistence diagrams are capable of capturing the complexity of the structure of the active regions in a very compact manner, allowing us not only to differentiate between different types of active regions but also to detect the formation of interesting structures, such as $\delta$-spots, structures in which umbrae of opposite polarities share a common penumbra, which are strongly associated with flare activity. 

\subsubsection{Active region classification}

It is important to understand which types of magnetic structures can be identified through persistent homology and establish the correspondence between these structures and the position of the corresponding topological feature in a persistence diagram. To achieve this understanding, Fig. \ref{fig:  AR - Complexity} displays both a magnetogram with complex morphology (panel b) and its corresponding persistence diagram (panel a), along with three zoomed-in regions of the magnetogram (panels c to e). In all panels, some topological features or their corresponding magnetic structures have been color-coded based on their types, or equivalently, based on their positions in the persistence diagram. In the complete magnetogram (panel b), structures have been marked with a cross, indicating the pixel where the structure died during the filtration process. Meanwhile, in panels c to e, all pixels composing each structure have been colored. It is noteworthy that nearly all pixels appear colored because we have selected the most significant structures—those with longer lifetimes (Eqn. \eqref{eq: lifetime}). Consequently, these structures encompass all less significant structures that are absorbed and incorporated into the former during the filtration process.

The analysis of the persistence diagram allows us to deduce several properties of the magnetogram. Firstly, the persistence diagram provides a rapid assessment of the intensity of the magnetic flux since the death value of the topological feature coincides with the maximum flux (in absolute value) within the corresponding structure. In the case of Figure 6, we observe that several structures exhibit maximum (absolute) values surpassing 1500 G, with multiple structures falling within the range of 1000 G to 1500 G.

Secondly, we can infer how the magnetic signals are distributed by examining the number of isolated and connected structures in the diagram (structures highlighted in blue and red depending on their polarity in Fig. \ref{fig:  AR - Complexity}). The complex morphology of the structure displayed in the magnetogram is evident in the high number of connected structures (regions II and V in the diagram) and the absence of prominent isolated structures (regions I and IV).

Lastly, the presence or absence of ring-like structures provides insights into how the magnetic structures are connected. These features can only be found in regions where connected structures create highly complex morphologies with gaps between them, as illustrated in the magnified regions of the magnetogram in Figure 6 (panels c to e).

\begin{figure}
    \centering
     \includegraphics[width=9cm]{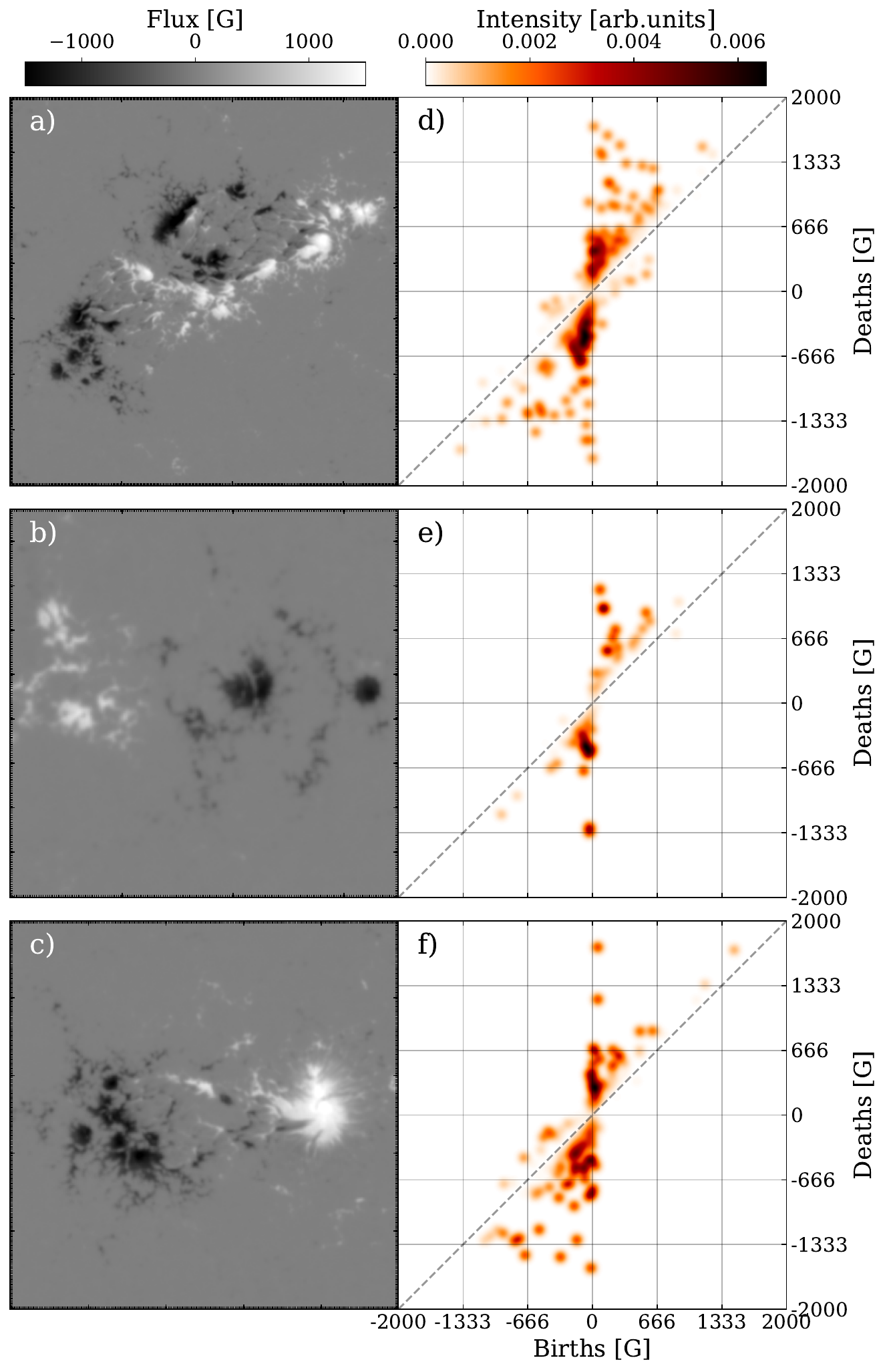}
    \caption{SDO/HMI magnetograms of three different active regions and their corresponding persistence images. a) NOAA AR 11158, date: 2011-02-13 at 06:34. b) NOAA AR 11098, date: 2010-08-12 at 20:58. c) NOAA AR 11072, date: 2010-05-22 at 20:58. All persistence images have been generated with the following parameters: Resolution =  1000 pixels$^2$ (4 G), weighting function: $\omega (\pi) = \arctan (5\times 10 ^{-8} \pi ^{3})$ and a gaussian kernel with $\sigma = 40\ G$.}
   \label{fig: AR-classification}
\end{figure}

An example of how the three features allow us to classify ARs depending on their morphologies is shown in Figure \ref{fig: AR-classification}, where three different ARs and their corresponding PIs are displayed. Although at first sight, the PIs appear to be very similar, especially the ones shown in panels d) and f), upon closer inspection, it is possible to find the differences when focusing on the three features mentioned previously. The first AR (panel a)), also shown in Fig. ~\ref{fig:  AR - Complexity}, shows a very complex morphology, where the magnetic field of both positive and negative polarities is distributed in multiple connected structures. This behavior is displayed in the PI through the high density in the isolated and connected features in equal proportions (i.e. with no prominent features) and with the presence of ring-like features in both polarities. In contrast, the second AR (panel b) shows a simpler magnetic structure with weaker signals. The PI for this case shows an absence of ring-like features in both polarities and a very low density in the regions for connected and isolated features. Lastly, panel c) shows an AR where the positive magnetic field is concentrated in one main area whereas the negative polarity magnetic field shows a fragmented structure. Comparing the corresponding PI with that of the initial case (panels f and d, respectively), a similarity is evident in the region corresponding to negative polarities, observed in both the density of connected components and the presence of ring-type structures. Nevertheless, when examining the positive polarity, it becomes evident that, unlike its negative counterpart, there are only a few prominent isolated structures and a notably low density of connected structures. Furthermore, this asymmetry between the positive and negative distributions is underscored by the absence of ring-like structures in the former.

\subsubsection{`Interacting' Diagram}
So far, we have shown how persistent homology is capable of identifying the various morphologies of active regions and the types of structures that can be identified through persistent images or diagrams. Nevertheless, these structures are either isolated or regions of the same polarity that interact with each other. Due to the nature of the filtration process, persistent homology is unable to detect structures where magnetic fields of opposite polarities form a joint structure. However, ARs where there is a significant interaction between magnetic fields of opposite polarities are of greater interest due to their association with flare production. 

This issue is illustrated in the analysis carried out in the previous section of NOAA Active Region 11158 (see Fig. ~\ref{fig:  AR - Complexity}). Although we can capture the complexity of both positive and negative magnetic structures through the persistent diagram, the $\delta$-spots present in the central region remain unnoticed. However, the large number of flare events associated with this region including an X2.2-class event are thought to be related to the abundance of these structures (e.g. \citealt{x21}, \citealt{x22}). In this section, we describe a way to efficiently detect and quantify these structures using persistence homology, with only a few additional steps in the analysis.

\begin{figure}
    \centering
     \resizebox{\hsize}{!}{\includegraphics[width=12cm]{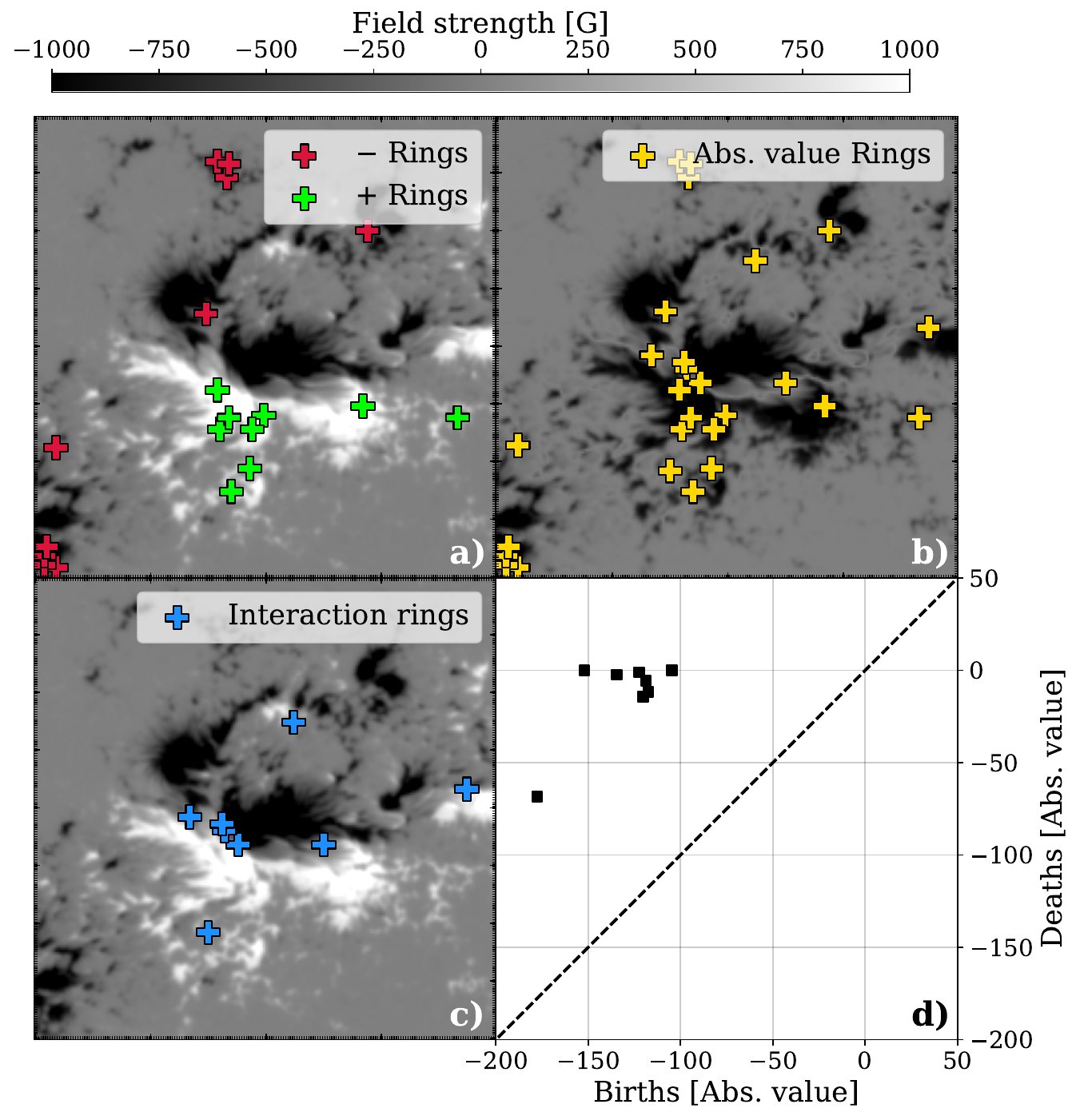}}
    \caption{Depiction of the different steps carried out for the generation of the interaction diagram (panel d)) for a zoomed-in region of an SDO/HMI magnetogram of NOAA AR 11158 observed on the 2011-02-13 at 18:34. The crosses point to the position at which each of the ring-like structures die, which is always located inside the perimeter of the hole. Panel a) shows the original magnetogram and the position where rings of positive and negative polarities are found. Panel b) shows the magnitude of the field strength in negative values along with the rings found in this representation. Panel c) shows the original magnetogram again, but now featuring the interaction rings, identified when comparing the position of the rings found in the previous steps. Lastly, panel d) shows the `interaction diagram' generated using only the interacting ring structures.}
   \label{fig: Interacting_diagram}
\end{figure}

We start by tracking the position in the magnetogram where rings are detected. Following this, we modify the magnetogram by inverting the sign of one polarity, ensuring that all pixels are either negative or positive. This way, we construct a second `magnetogram' in which we only have information about the intensity, in absolute value, of the magnetic field. This allows us to identify features formed by any combination of signals, whether they are of the same polarity or opposite. Using this new magnetogram, we repeat the analysis and record the positions in the magnetograms at which we find ring-like features. It is worth noting that in this analysis, all the rings identified in the initial step, which were formed by structures of equal polarities, are still detected, although their position in the persistent diagram may have changed due to the change in polarity, hence the relevance of tracking the pixel in the magnetograms. However, only in this second analysis can we identify rings formed between opposite polarities. By selectively considering the ring-like features exclusively identified in this second analysis, we can effectively identify and characterize the structures that are formed by the interaction between different polarities. 

Figure \ref{fig: Interacting_diagram} depicts an example of the interaction analysis for NOAA AR 11158, which exhibits a strong interaction between opposite-polarity fields. Panel a) displays the magnetogram, indicating also the positions where ring-like features have been identified for both positive and negative polarities. Only features with absolute persistencies greater than $5\sigma _ {bg}$ and whose birth occurs in the range: ($5\sigma _ {bg}, \infty$), for positive polarities, and: ($-\infty, -5\sigma _ {bg}$) for negative polarities, have been recorded. We now repeat the same analysis but using only the magnitude of the signal. To do this, we invert the positive polarity and, once again, register the positions of the ring-like features in this new magnetogram, as shown in panel b). As observed in panel c), the majority of the interaction rings (\textit{i.e.} those exclusively identified in the second analysis) are located in areas characterized by strong interaction, where the $\delta$-spots were found. In these areas, both polarities interact, resulting in the formation of ring-like structures comprising positive and negative magnetic fields due to their close proximity. These areas, such as  $\delta$-spots, are of particular interest, as they typically harbor magnetic cancellation, reconnection, and flux emergence.  However, some points appear to be situated in uni-polar fields. These points, despite what may appear at first sight, are found by this analysis due to a structure that requires the other polarity to close completely and thus form a ring. It is noteworthy that the occurrences of such cases are quite limited when compared to the rings observed in highly interacting zones. While their presence does not necessarily indicate intense interaction, it does imply a certain level of interaction between the two polarities. These features can be represented in a persistence diagram in an analogous way to the standard results of a persistent homology analysis. This is what we have referred to as `interacting diagram' and it is worth noting that only the magnitude of the birth and death coordinates are relevant parameters since the sign will be the one matching the polarity selected at the second step. We have chosen to invert the positive polarity so that these features have positive persistencies, as it is more common in persistent homology studies. However, it is important to emphasize that this decision is completely arbitrary and has no impact on the results of the analysis.  

It is useful to determine the information conveyed by the interaction diagram regarding the structures themselves. By taking into account the position and quantity of the interaction rings, along with the temporal evolution of the diagram, we can discern the moment and location where these highly interacting structures develop. Therefore, interaction diagrams could be a new tool to identify, through their topological properties, the strong-gradient polarity inversion lines that characterize $\delta$-spots. To achieve this, it is necessary to incorporate into the analysis the temporal evolution of these structures and study the properties that can be extracted from the $\delta$-spots through these diagrams, which goes beyond the scope of this work but represents the next (necessary) step to asses the predicting capabilities of persistent homology in the field of solar physics.

\section{\label{sec : Conclusion}Conclusion}

In this study, we investigate the most adequate approach for the application of persistent homology algorithms to the analysis of solar magnetograms. By combining different filtrations in a single one-dimensional persistent homology analysis, we can effectively capture structures corresponding to both polarities of the magnetic field. We have applied this analysis to observations of the quiet Sun and active regions, taken with both Hinode/SOT and SDO/HMI, respectively. Lastly, we have analyzed the results and identified the features of the data that can be found through persistent diagrams and images, and also show some examples of applications of the algorithms. 

Our proposed approach to persistent homology algorithms involves the integration of sublevel and superlevel filtrations within a single analysis, enabling the creation of a comprehensive persistence diagram that encompasses features from both positive and negative magnetic structures. Through the examination of the positions of these identified features within the resulting persistence diagram, we can discern the diverse magnetic features present in the magnetograms. This approach has demonstrated its efficacy in capturing the intricate complexity of magnetic structures, with a particular emphasis on active regions. Through this method, we have achieved successful differentiation between the various morphologies present in active regions by analyzing the presence or absence of specific features in the corresponding persistence images. 

On the other hand, the persistent images obtained from quiet Sun observations exhibit significant similarity to each other. This indicates a lack of overall evolution in the magnetic structures within these regions. In quiet Sun areas, small regions of magnetic flux interact with each other in small-scale events, while the overall structure remains relatively static. These small-scale events become more apparent in persistent images when the field of view is reduced. These small-scale events, such as flux emergence or cancellation, can be observed through persistent images as a joint movement of negative and positive features. In cancellation events, the features move toward the center of the image, while in emergence events, they move away from the center.

Additionally, we have successfully identified interactions between opposite-polarity magnetic fields by detecting ring-like features formed by these two polarities. To achieve this, we introduced a method for calculating an `interaction diagram' that selectively displays features resulting from the interaction between polarities. This interaction diagram is generated by comparing the ring-like features identified in an analysis using only the absolute value of the signal with those found in the standard analysis. This approach enables us to detect the presence of $\delta$-spots and quantify the level of interaction between polarities, which is one of the critical factors for the understanding and prediction of flare eruptions. 

In conclusion, our application of persistent homology to solar magnetograms has provided a comprehensive and insightful framework for studying magnetic structures on the solar surface. The topological features derived from magnetograms serve as a foundation for classifying active regions based on their morphology and level of interaction, as certain topological features may have inherent connections to solar atmospheric phenomena. For instance, the presence of interaction rings in active regions might be correlated with flare production, while the interaction of signals from opposite polarities observed in a persistent diagram in the quiet Sun could be linked to small-scale reconnection events or the separation of signals associated with flux emergence. The exploration of these relationships and the assessment of the presented tools in achieving precise active region classification and their potential as predictive tools are topics of our upcoming research. Moreover, we have introduced new tools, such as the interaction diagram, which facilitates the detection and quantification of structures interacting with opposite polarities, like $\delta$-spots, addressing a crucial aspect of flare prediction. The findings presented in this article lay a solid foundation for future studies, emphasizing the potential of persistence images as valuable inputs for machine learning algorithms and contributing to advancements in space weather forecasting.

Lastly, it is important to emphasize that in this study we have focused primarily on static images in order to provide a solid basis for future investigations. The next logical step in this study is to complete the analysis of active regions, which includes examining their temporal evolution. This approach allows for the simultaneous consideration of two key factors in understanding flare eruption processes: morphological complexity, whose analysis is intrinsic to persistent homology, and the study of their temporal evolution through the analysis of the evolution of persistence and interaction diagrams.
\pagebreak
\section*{Acknowledgements}
We are grateful to the Japan Aerospace Exploration Agency's Institute of Space and Astronomical Science (ISAS/JAXA) and the SDO team for the distribution of Hinode's and HMI data. P.S.G. thanks the authors of the Homcloud library for freely publishing the code. This work was supported by AEI/MCIN/10.13039/501100011033/ (RTI2018-096886-C5, PID2021-125325OB-C51, PCI2022135009-2), ERDF “A way of making Europe”;  JSPS KAKENHI grant numbers JP18H05234, JP23H01220 (PI: Y. Katsukawa), NAOJ-RCC-2201-0402 and -0403;  and by JSPS KAKENHI grant Nos. JP20KK0072 (PI: S. Toriumi), JP21H01124 (PI: T. Yokoyama), and JP21H04492 (PI: K. Kusano).

\vspace{5mm}

\software{Homcloud \citep{Homcloud}}

\bibliography{Bibliography}{}
\bibliographystyle{aasjournal}

\end{document}